\documentclass[12pt,dvips]{article}
\include{epsf}
\begin{document}
%%%%%%%%%%%%%%%%%%%%%%%%%%%%%%%%%%%%%%%%%%%%%%%%%%%%%%%%%%%%%%%%%%%%%%%%%%%
\thispagestyle{empty}
\begin{flushright}
UNIGRAZ-UTP-09-03-99 \\
MIT-CTP-2899 \\
hep-lat/9909025
\end{flushright}
\begin{center}
\vspace*{10mm}
{\Large The chiral limit of the two-flavor lattice 
\vskip2mmSchwinger 
model with Wilson fermions$^*$}
\vskip11mm
{\bf Christof Gattringer}
\vskip2mm
Massachusetts Institute of Technology \\
Center for Theoretical Physics \\
77 Massachusetts Avenue, Cambridge MA 02139 USA
\vskip6mm
{\bf Ivan Hip and C.~B. Lang}
\vskip2mm
Institut f\"ur Theoretische Physik \\
Universit\"at Graz, A-8010 Graz, Austria
\vskip14mm
\begin{abstract}
We study the 2-flavor lattice Schwinger model with Wilson fermions in the 
chiral limit. The quark mass is determined using the PCAC definition. 
We numerically compute the masses of the iso-triplet ($\pi$) and iso-singlet 
particles ($\eta$) for different quark masses and compare our 
results with analytical 
formulas.
\end{abstract}
\end{center}
\vskip8mm
\noindent
PACS: 11.15.Ha \\
Key words: Lattice field theory, Schwinger model. \\
To appear in Physics Letters B.
\vskip7mm \nopagebreak \begin{flushleft} \rule{2 in}{0.03cm}
\\ {\footnotesize \ 
${}^*$ Supported by Fonds zur F\"orderung der Wissenschaftlichen 
Forschung in \"Osterreich, Project P11502-PHY}
\end{flushleft}
\newpage
\setcounter{page}{1}

\subsection*{Introductory remarks}
Although we begin to understand the formulation of chiral fermions on the 
lattice \cite{chiral}, the high cost of their simulation  still renders
Wilson fermions an attractive alternative. Wilson fermions, however, suffer
from an  explicit breaking of chiral symmetry and can  only in a carefully
taken chiral limit be used to analyze chiral aspects of QCD. Approaching and
properly understanding this limit is a non-trivial  task and studies in
2-dimensional models contribute to our understanding.

The Schwinger model \cite{Sch62} with two flavors is a particularly convenient
test-bed  for studying the chiral limit since it shares several  characteristic
features with QCD.  The low lying spectrum (``mesons'')  contains an
iso-triplet of light particles ($\pi$'s)  which become massless in the chiral
limit. In addition a massive iso-singlet state ($\eta$) appears, giving rise to
a U(1) problem equivalent to QCD. For the massless case the analytical solution
of the Schwinger model in the  continuum has been known for many years and
also for small fermion masses some analytical results for the model with flavor
have become available  \cite{CoJaSu75}-\cite{Sm97}. 

For the lattice model with staggered fermions the connection to these 
continuum results is well understood \cite{MaPaRe81}-\cite{BeGrSeSo99}. 
The challenge in a lattice simulation with Wilson fermions is to properly 
fine-tune the bare parameters and drive the pions to zero mass. Although  many
aspects of the Schwinger model with Wilson fermions are well  understood - in
particular the relation between the  spectrum of the Dirac operator and  the
topology of gauge fields \cite{Vi88}-\cite{HoLaTe98} has been studied
intensively - relatively little \cite{IrSe96}-\cite{ElBu97}, \cite{HiLaTe98} is
known for the  chiral limit with Wilson fermions. What is still lacking is a
high precision  numerical study of the mass spectrum for Wilson fermions
combined  with a systematic comparison to analytical results.  Furthermore the
Schwinger  model allows for a simple implementation of the  PCAC method
\cite{pcac} for defining the  quark mass. Using this definition together with
the numerical results  for the mesons in the comparison to analytical formulas
leads to an  interesting test of the PCAC quark mass definition.

In this letter we report on our study of  the two-flavor lattice Schwinger
model  with Wilson fermions. In particular we concentrate on the low-lying
particle  spectrum ($\pi$'s, $\eta$) and the behavior in the chiral limit. We
use the PCAC method to define the quark mass $m$ and compare our results for
the spectrum of the meson masses as a function of $m$ with analytical formulas.

\subsection*{Setting, quark mass and observables}

The action for the Schwinger model with Wilson fermions consists of two parts
$S = S_G + S_F$ with $S_G$ and $S_F$ being the  gauge field and fermion action
(2 flavors) both in standard Wilson  formulation. We work on $L \times L$
lattices with periodic boundary  conditions for the gauge fields and mixed
periodic boundary conditions (periodic in space-, anti-periodic in
time-direction) for the fermions. We simulate the system on lattices of size
$L=16, L= 24$ and $L=32$ and  $\beta$-values of $\beta = 2,4,6$. For the
smallest lattice we use $10^4$ configurations for the determination of the
observables, while for $L=24$ and $L=32$ we typically use $10^3$ 
configurations but increase this number for some values of the parameters where
better statistics is desirable. The update is done using the standard hybrid
Monte Carlo method \cite{DuKePeRo87} with 10-steps trajectories and a step size
adjusted such that the average acceptance rate is 0.8. Several observables,
including the geometric  topological charge, are monitored and appropriate
numbers of configurations are discarded between the measurements to assure
proper de-correlation and achieve topological ergodicity.

The quark mass is determined using the PCAC method as outlined in  \cite{pcac}.
The essential idea is to use the coefficient of the  right hand side of the
axial SU(2) Ward identity as a definition of the mass. One computes the matrix
elements with a pseudo-scalar 
source on both sides  of the Ward identity and the
quark mass is then given by the ratio of the vev.'s for the two sides. The
details on the implementation of the method for the  lattice Schwinger model
were reported elsewhere \cite{HiLaTe98}.  Here we just briefly summarize the
relevant formulas. The quark mass is  defined as ($x \neq y$, $\mu=1,2$)
\begin{equation}
m \; = \; \frac{1}{2} \frac{ \sum_\mu \Big[ 
\langle 0 | \pi_y^{a \dagger} J_x^{5a\mu} | 0 \rangle - 
\langle 0 | \pi_y^{a \dagger} J_{x-\mu}^{5a\mu} | 0 \rangle \Big]}{
\langle 0 | \pi_y^{a \dagger} \pi_x^a | 0 \rangle} \; ,
\label{pcacmass}
\end{equation}
with
\begin{eqnarray}
\pi^a_x & = & \frac{1}{2} \overline{\psi}_x \; \tau^a \otimes 
\gamma_5 \; \psi_x \; , \nonumber \\ 
J_x^{5a\mu} & = & \frac{1}{4} \Big[ 
\overline{\psi}_{x+\hat{\mu}} \,U^\dagger_{x,\mu}\,\tau^a \otimes \; 
\gamma_\mu \gamma_5 \; \psi_x
\; + \; \overline{\psi}_x \; \tau^a \otimes \gamma_\mu \gamma_5 \,U_{x,\mu}\,
\psi_{x+\hat{\mu}} \Big] \; .
\end{eqnarray}
The gamma matrices are chosen as $\gamma_1 = \sigma_1, \gamma_2 = \sigma_2$
and $\gamma_5 = \sigma_3$ with $\sigma_j$ denoting the Pauli matrices. 
The generators $\tau^a$ of rotations in flavor space are also given by 
the Pauli matrices $\tau^a = \sigma_a, \; a = 1,2,3$. The quark mass $m$ was 
computed for each $\beta$, $L$ and $\kappa$ individually.

The masses for the pions and the eta-particle were computed from the 
decay of two point functions of the following {\em currents}
\begin{eqnarray}
J_x^{a \mu} & = & \overline{\psi}_x \; \tau^a \otimes \gamma_\mu \; \psi_x
\; \; \; \; \; \mbox{for the $\pi$'s} \; , 
\nonumber \\
J_x^{0 \mu} & = & \overline{\psi}_x \; \mbox{1\hspace{-1.4mm}1} 
\otimes \gamma_\mu \; \psi_x \; \; \; \; \; \; \; \mbox{for the} \;
\eta \; .
\label{currents}
\end{eqnarray} 
We emphasize that 2-point functions of {\em scalars} or  {\em pseudo-scalars}
are not  particularly well suited for the determination of the meson masses.
These operators are bosonized (compare the discussion below) by cosines and 
sines of the fundamental fields \cite{BeSwRoSch79,GaSe94} and their 2-point 
functions strongly mix contributions from both the triplet and singlet states.

Finally let us briefly discuss the correspondence of the gauge coupling on the
lattice to the gauge coupling $g$ as it appears in the continuum formulas 
below. The Schwinger model is a super-renormalizable theory, i.e.~the  bare
coupling does not get renormalized and equals the physical coupling $g$.
 
On the lattice we use the usual factor $\beta$ in front of the gauge field
action. In the naive  continuum limit it is related to the continuum coupling
$g$ via (we set the lattice spacing $a = 1$) \begin{equation}
\beta \; = \; \frac{1}{g^2} \; .
\label{couplingrelation}
\end{equation}
Again we can use the super-renormalizability of the model and define the
coupling $g$ which we use for comparison with the analytical 
results in the continuum (see below) through (\ref{couplingrelation}). 

\subsection*{Analytical results for the continuum model}

Before we start with the presentation of our numerical results, let us briefly 
discuss the known analytical results for the continuum model. This is best
done  by looking at the generalized Sine Gordon model which bosonizes the
2-flavor Schwinger model \cite{Co76}. Its Lagrangian reads
\begin{equation}
\frac{1}{2}\Big(\partial \varphi^{(0)}\Big)^2 +
\frac{1}{2}\Big(\partial \varphi^{(3)}\Big)^2 +
\frac{1}{2}\, \frac{2\,g^2}{\pi} \Big(\varphi^{(0)}\Big)^2 
- \frac{m \,c}{\pi} \cos\Big(\sqrt{2 \pi} \,\varphi^{(0)}\Big) 
\cos\Big(\sqrt{2 \pi} \,\varphi^{(3)}\Big) \; .
\label{gensg}
\end{equation}
The constant $c$ is not determined by the bosonization but is related  (see
e.g.~\cite{GaSe94,Ga96}) to the masses $\mu^{(0)}$ and $\mu^{(3)}$ used for
normal ordering of the fields  $\varphi^{(0)}$ and $\varphi^{(3)}$ by  $c =
(\mu^{(0)} \mu^{(3)})^{1/2} e^\gamma /2$, where $\gamma$ denotes Euler's
constant $\gamma = 0.577216\ldots \; $. For the case of two flavors two of
the currents which we measure (compare (\ref{currents})) are bosonized with the
following prescription (here $\tau^0 \equiv$ 1\hspace{-1.4mm}1)
\begin{equation}
\overline{\psi}(x) \; \tau^a \otimes \gamma_\mu \; \psi(x)
\; = \; \frac{1}{\sqrt{\pi}}\, \varepsilon_{\mu \nu}\, \partial_\nu \;
\varphi^{(a)}(x) \quad , \qquad a = 0,3 \; .
\label{bosonization}
\end{equation}
For the other two members $j_\mu^{(1)}, j_\mu^{(2)}$ of the iso-triplet no
abelian bosonization is known. However since the original model  is invariant
under flavor rotations, their masses are the same as for the triplet current
($a=3$, pions). When setting the quark mass $m$  in (\ref{gensg}) to zero, the
two flavor Schwinger model is bosonized by two free fields. One of them, the
pion field $\varphi^{(3)}$, is massless ($M_\pi = 0$) and the eta field 
$\varphi^{(0)}$ obtains the Schwinger mass $M_\eta = g \sqrt{2/\pi}$.

For non-vanishing quark mass, also the bosonized model (\ref{gensg}) can no
longer be solved in closed form. A first approach is to use a  semi-classical
analysis  (see e.g.~\cite{Ga96}) which gives a good approximation when all
involved masses are large. One simply determines the minimum of  the
interaction part $V[\varphi^{(0)},\varphi^{(3)}]$ of (\ref{gensg})  and the
squares of the masses are then  given by the second derivatives of
$V[\varphi^{(0)},\varphi^{(3)}]$  at the minimum.  After normal ordering the 
fields $\varphi^{(a)}$ with respect to their own masses (i.e.~setting 
$\mu^{(0)} = M_\eta$ and $\mu^{(3)} = M_\pi$), the semi-classical analysis for
the iso-triplet and iso-singlet gives
\begin{equation}
\frac{M_\pi}{g} \; =  \; e^{2\gamma/3}\; \frac{2^{5/6}}{\pi^{1/6}} \;  
\left(\frac{m}{g}\right)^{2/3} \; 
= \; 2.1633\ldots \; \left(\frac{m}{g}\right)^{2/3} \; ,
\label{semitrip}
\end{equation}
and
\begin{equation}
\frac{M_\eta}{g} \; = \; \sqrt{\frac{2}{\pi} \; + \; 
\Big(\frac{M_\pi}{g}\Big)^2}
\; .
\label{semising}
\end{equation}
We remark that the latter equation, i.e.~the semi-classical relation between 
$M_\eta$ and $M_\pi$, holds independently of any normal ordering prescription 
or the value of $M_\pi$, due to the formula $\partial^2/\partial
{\varphi^{(0)}}^2 V =  g^2 2/\pi + \partial^2/\partial {\varphi^{(3)}}^2 V$,
which in turn is evident  from (\ref{gensg}). This will be explored below. 

In an attempt to go beyond semi-classical analysis one can approximate  the
generalized Sine-Gordon model (\ref{gensg}) by a solvable model. One
possibility \cite{Co76} is to analyze the model in the  limit of large coupling
$g$ and small mass $m$ where the iso-singlet field $\varphi^{(0)}$ becomes
static and the model is reduced to a standard  Sine-Gordon model for the
triplet field $\varphi^{(3)}$
\begin{equation}
\frac{1}{2}\Big(\partial \varphi^{(3)}\Big)^2 
\; - \; 2C \cos\Big(\sqrt{2 \pi} \varphi^{(3)}\Big) \; .
\label{standardsg}
\end{equation}
This reduced bosonized theory has been studied \cite{Co76,Gr91} using the WKB
approximation \cite{DaHaNe75} for (\ref{standardsg}). The spectrum of  the
model (\ref{standardsg}) contains a soliton, an anti-soliton and two
soliton--anti-soliton bound-states. The lighter bound state has a mass equal 
to the soliton (and anti-soliton), and the mass for this triplet can be related
to the mass of the iso-triplet in the 2-flavor Schwinger model. More recently
(\ref{standardsg}) was studied again by  Smilga \cite{Sm97} but his analysis is
now based on the newly found  analytic solution  \cite{Za95} of the standard
Sine-Gordon model. Smilga finds 
\begin{equation}
\frac{M_\pi}{g} \; = \; 2^{5/6}\, e^{\gamma/3} \,
\left( \frac{\Gamma(\frac{3}{4})}{\Gamma(\frac{1}{4})} \right)^{2/3}
\frac{\Gamma(\frac{1}{6})}{\Gamma(\frac{2}{3})} \; 
\left(\frac{m}{g}\right)^{2/3} \; = \;
2.008\ldots \; \left(\frac{m}{g}\right)^{2/3}\; .
\label{smilgatrip}
\end{equation} 
Obviously the truncation of the original model (\ref{gensg}) to the 
standard form (\ref{standardsg}) does not allow for a result for the 
singlet mass $M_\eta$. For this state only the semi-classical 
formula (\ref{semising})
is available. It his however interesting to use the exact result
(\ref{smilgatrip}) of the 
truncated model as an input in (\ref{semising}), and below
we compare also this formula to the numerical data.

\subsection*{Numerical results}

\begin{figure}[htpb]
\centerline{
\epsfysize=5.8cm \epsfbox[ 31 10 525 440 ] {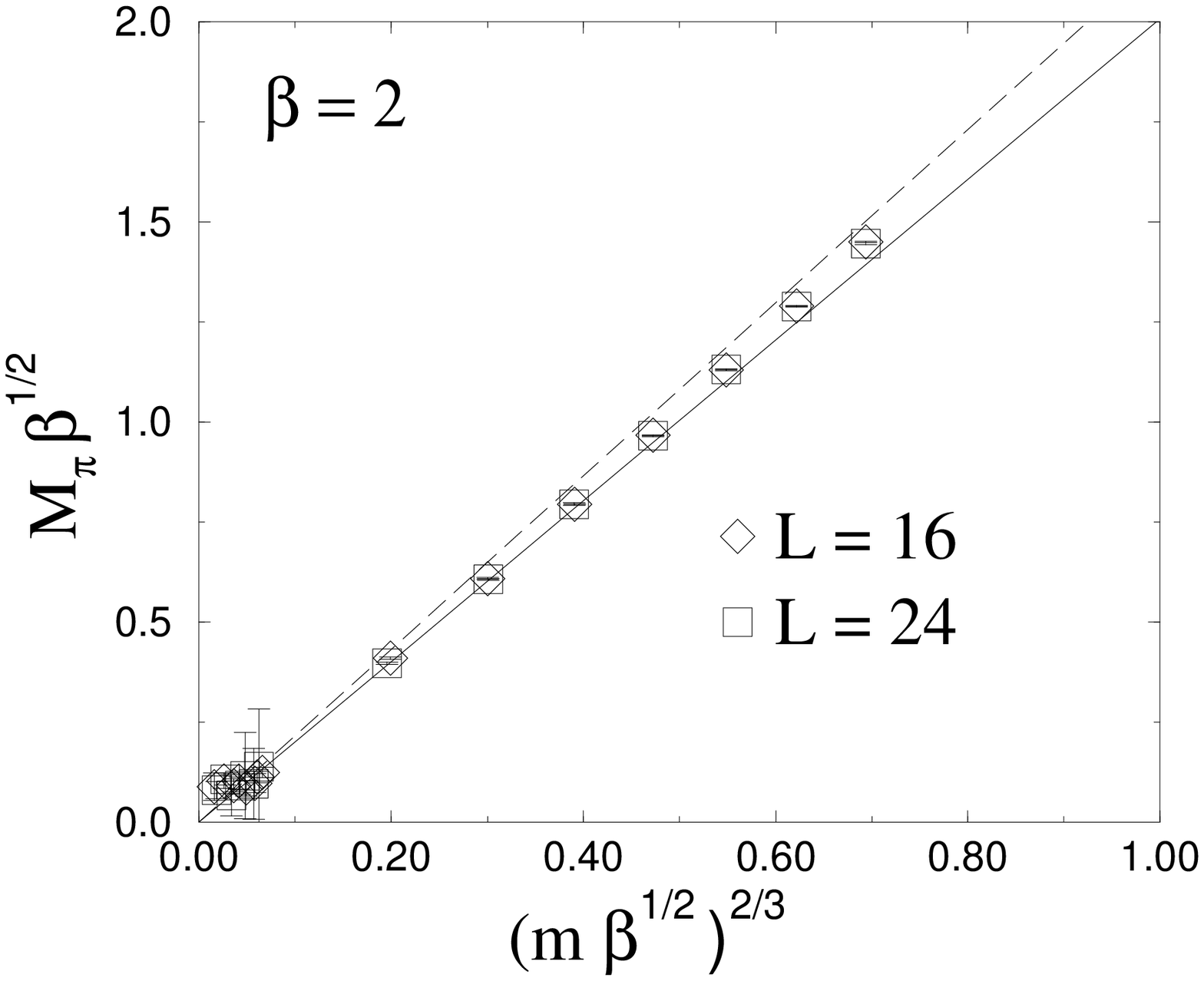}
\epsfysize=5.8cm \epsfbox[ 67 10 525 440 ] {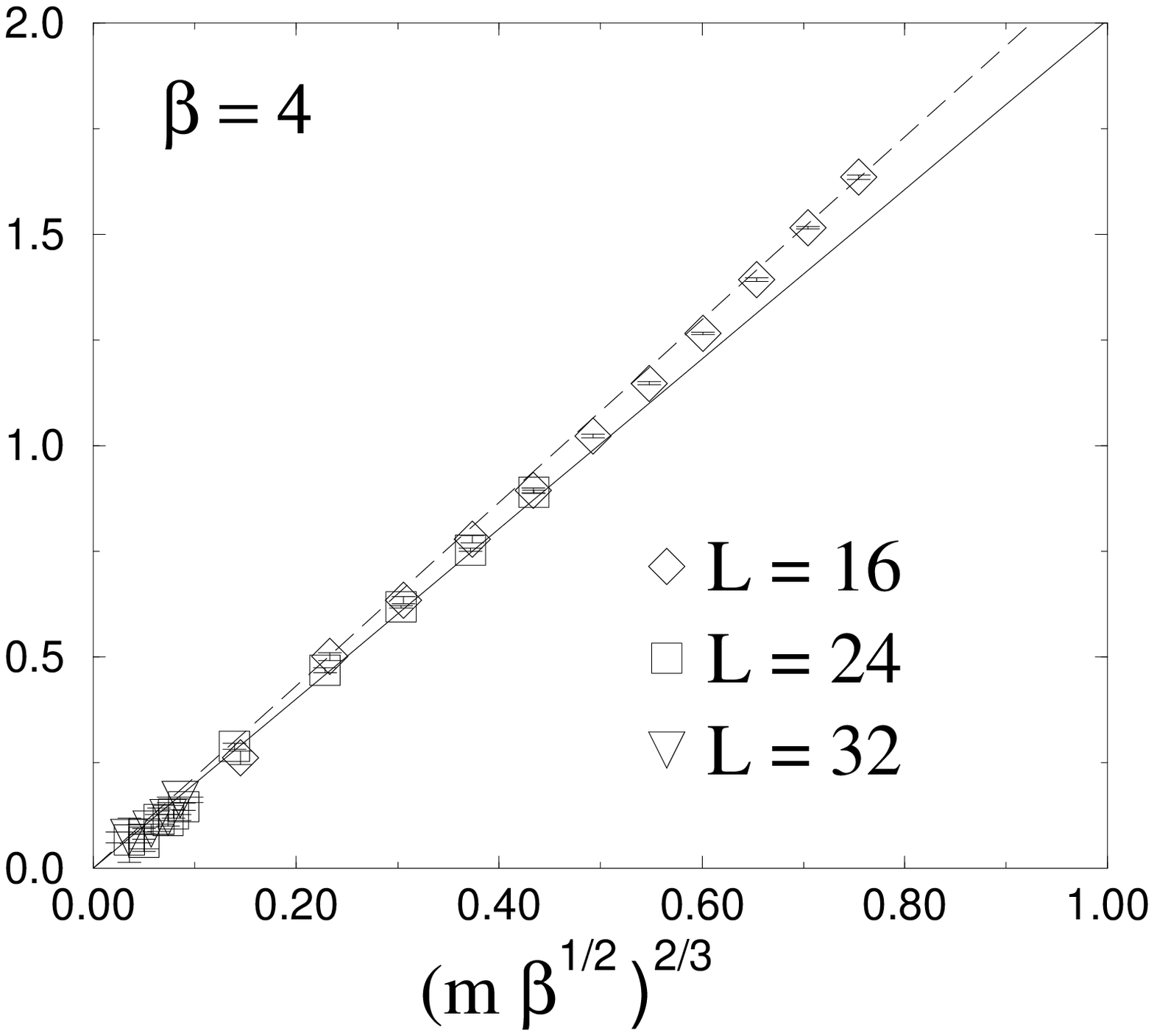}}
\vspace{2mm}
\centerline{
\epsfysize=5.8cm \epsfbox[ 31 10 525 440 ] {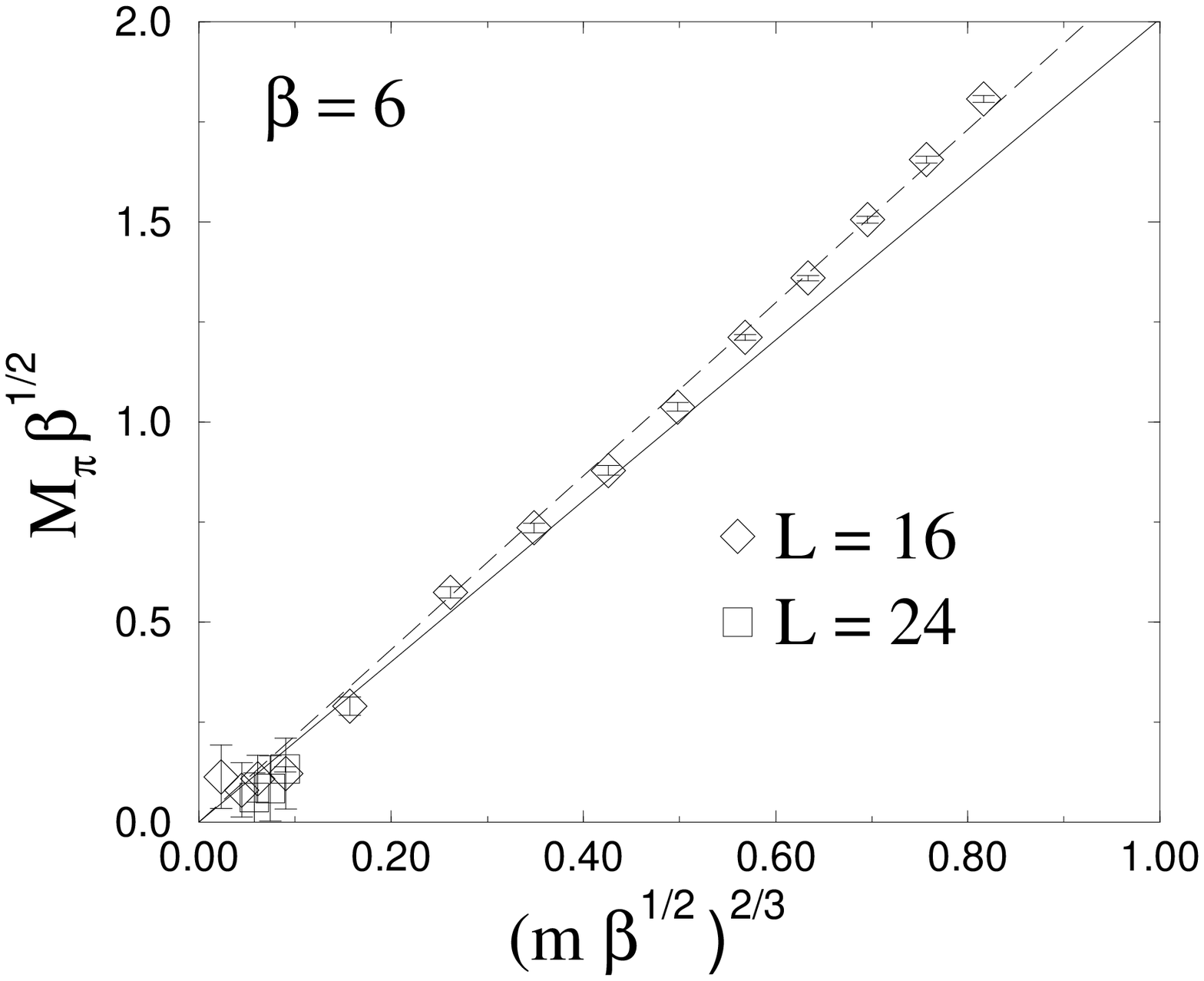}
\epsfysize=5.8cm \epsfbox[ 67 10 525 440 ] {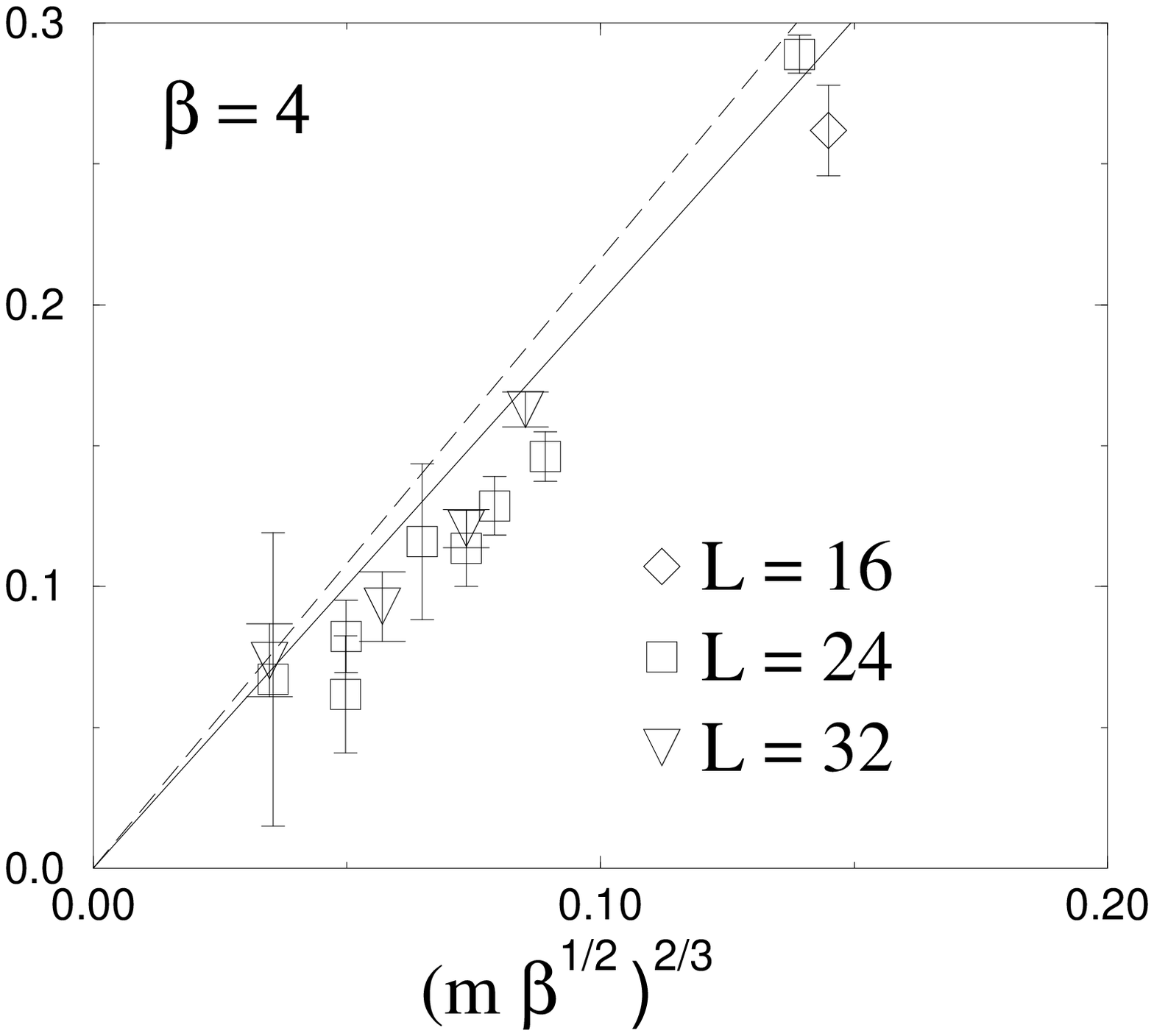}}
\caption{{\sl Results for the pion mass $M_{\pi}$. 
Symbols: Monte Carlo data. Full line: Smilga's formula} 
(\protect{\ref{smilgatrip}}). {\sl Dashed line:
the semi-classical formula} 
(\protect{\ref{semitrip}}).
\label{tripplot}}
\end{figure}

Let us now discuss our numerical results for the mass spectrum.  We start with
presenting the data for the triplet mass in Fig.~\ref{tripplot} and compare
them  to the semi-classical-result (\ref{semitrip}) and Smilga's formula
(\ref{smilgatrip}).  We show our results for $\beta = 2,4,6$ in the first three
plots and  and in each plot combine all available values for the sizes $L$. In
the last plot we zoom in on the chiral region of the $\beta = 4$ plot.  We
compare the numerical results to Smilga's formula (\ref{smilgatrip}) (full
line) and the semi-classical result (\ref{semitrip}) (dotted line). We remark,
that the errors in the determination of $m$ from (\ref{pcacmass}) are smaller
than the symbols and thus no horizontal error bars were  included in the plots.

\begin{figure}[htpb]
\centerline{
\epsfysize=5.8cm \epsfbox[ 31 20 525 438 ] {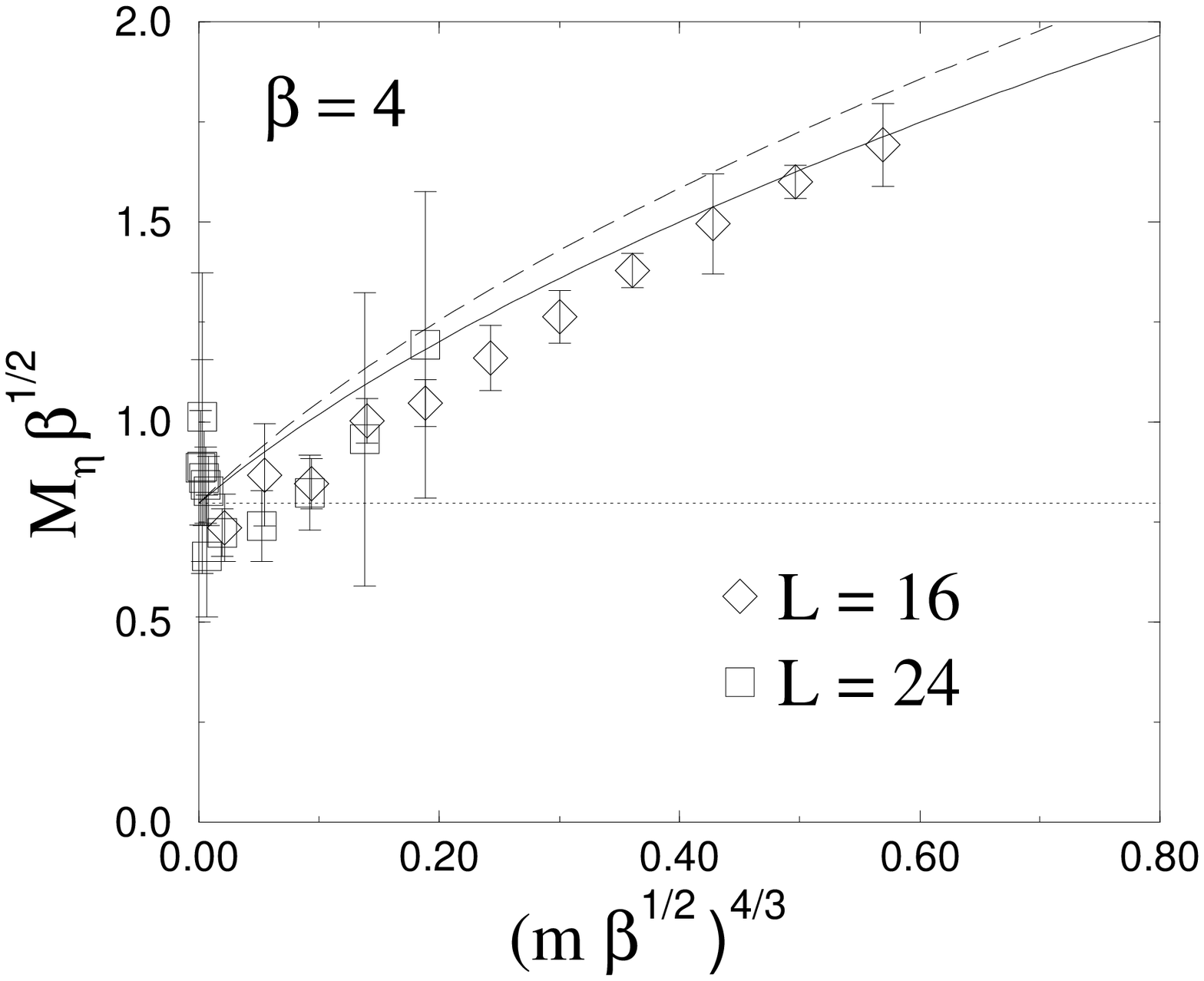}
\epsfysize=5.8cm \epsfbox[ 67 20 525 440 ] {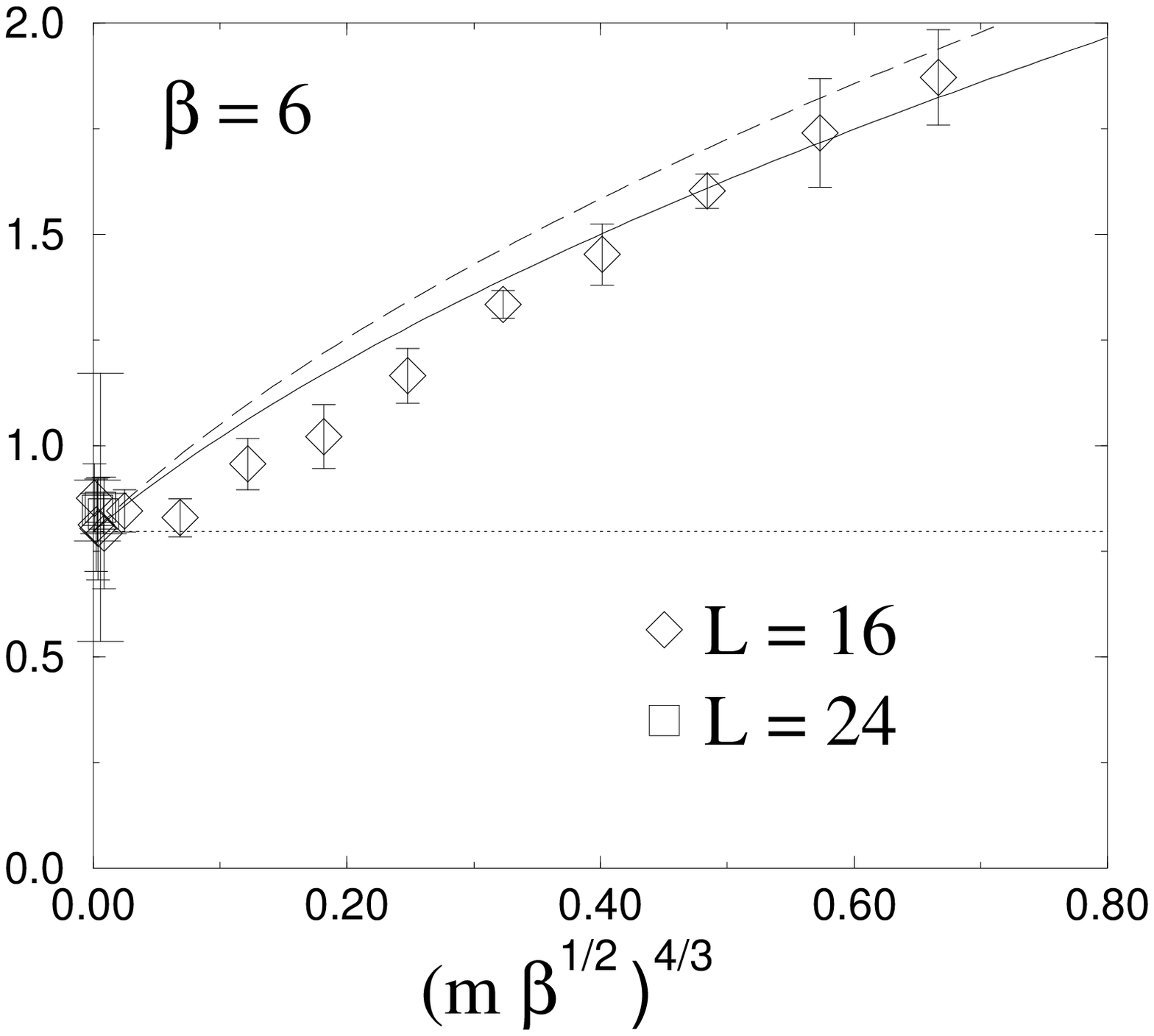}}
\caption{ {\sl Results for the eta mass $M_{\eta}$. Symbols: Monte 
Carlo data. Dashed curve: the semi-classical formula} 
(\protect{\ref{semising}}) {\sl with $M_\pi$ taken from}
(\protect{\ref{semitrip}}). {\sl Full curve: the semi-classical formula} 
(\protect{\ref{semising}}) {\sl with $M_\pi$ taken from}
(\protect{\ref{smilgatrip}}). {\sl The dotted horizontal 
line corresponds to $M_\eta$ at quark mass $m = 0$}. 
\label{singplot}}
\end{figure}

It is obvious, that the approach to the chiral limit is under good  control,
i.e.~the pion mass $M_\pi$ vanishes as the PCAC quark mass  goes to zero. In
addition the data is well described by the analytical result 
(\ref{smilgatrip}) (full line). In particular at small quark mass  and low
$\beta$ where Smilga's assumption $m \ll g$ is fulfilled best, the coincidence 
of the data with (\ref{smilgatrip}) is convincing.  For increasing quark  mass
we observe a cross-over behavior and the data is then better described  by the
semi-classical result (\ref{semitrip}) which is  represented by the dashed
line. This deviation from  (\ref{smilgatrip}) was also observed for staggered
fermions in    \cite{GuKaStWe98}. The  comparison of the data at different
lattice sizes $L$ shows, that the data points essentially fall on top of each
other and for our parameters  $L$ and $\beta$ the finite size effects are
negligible. When zooming in on the very small $m$ region (last plot in
Fig.~\ref{tripplot}) the error bars become larger due to the small  $M_\pi$,
but the approach to the chiral limit is still relatively clean.
\vskip2mm
In Fig.~\ref{singplot} we show our results for the eta mass $M_\eta$ at $\beta
= 4$ and $6$. We find that the data approaches the correct $m=0$ value (dotted
line) as the quark mass vanishes. The approach is  smoother as one goes closer
to the continuum limit, i.e.~increases $\beta$.  The semi-classical formula
(\ref{semising}) provides a reasonable  description of the data for larger $m$,
in particular when Smilga's result  (\ref{smilgatrip}) for $M_\pi$ is used as
an input in (\ref{semising})  (full curve). 

For small $m$ the quantum fluctuations become larger  and thus the numerical
data deviates from the semi-classical curve.  This behavior at small $m$ can be
fitted by a power law as has been done  in \cite{GuKaStWe98} for the results
from the model with staggered fermions.
\vskip2mm
To summarize, we find that the approach to the chiral limit is under good 
control. As the PCAC quark mass $m$ goes to zero we find that the pion mass
$M_\pi$ vanishes and $M_\eta$ approaches the Schwinger value $g \sqrt{2/\pi}$.
The data for $M_\pi$ is well described by Smilga's formula  (\ref{smilgatrip})
for $m \ll g$ with a cross-over to the semi-classical  formula (\ref{semitrip})
as $m$ is increased. For the eta mass $M_\eta$  the data is reasonably well
described by the semi-classical formula  (\ref{semising}) with quantum
fluctuations becoming less important as  the quarks are made heavier.   

{\bf Acknowledgement:}
We want to thank Andrei Smilga and Renate Teppner for discussions.

\end{document}